\documentclass[12pt]{amsart}
\usepackage{amsmath,amssymb,amsthm}
\newtheorem{proposition}{Proposition}

\newtheorem{corollary}{Corollary}

\theoremstyle{definition}
\newtheorem{remark}{Remark}
\newtheorem{example}{Example}
\newtheorem{definition}{Definition}

\newcommand{\N}{\mathbb N}

\newcommand{\R}{\mathbb R}

\newcommand{\ob}{\mathcal{O}}

\newcommand{\obso}{\mathcal{O}(\Omega,\hi)}
\newcommand{\obsoa}{\mathcal{O}(\Omega,\A,\hi)}

\newcommand{\loc}{\mathcal{O_L}}

\newcommand{\fsim}{\stackrel{f}{\sim}}
\newcommand{\csim}{\stackrel{c}{\sim}}  
\newcommand{\dsim}{\stackrel{d}{\sim}} 
\newcommand{\isim}{\stackrel{i}{\sim}}
\newcommand{\de}{\preccurlyeq_d} 
\newcommand{\coar}{\preccurlyeq_c}
\newcommand{\fuz}{\preccurlyeq_f} 
\newcommand{\info}{\preccurlyeq_i}
\newcommand{\hi}{\mathcal H}
\newcommand{\lh}{\mathcal{L(H)}}

\newcommand{\sh}{\mathcal{S(H)}}

\newcommand{\eh}{\mathcal{E(H)}}

\newcommand{\ip}[2]{\langle {#1}|{#2}\rangle} %innerproduct
\newcommand{\tr}{\mathrm{tr}}
\newcommand{\no}[1]{||{#1}||} %norm
\newcommand{\ket}[1]{|{#1}\rangle}
\newcommand{\bra}[1]{\langle {#1}|}
\newcommand{\ketbra}[2]{|{#1}\rangle\langle {#2}|}

\newcommand{\ran}[1]{\textrm{ran}({#1})}

\newcommand{\A}{\mathcal{A}}

\newcommand{\br}{\mathcal{B}(\R)}

\newcommand{\tn}{M^+_1(\Omega,\A)}
\newcommand{\tno}{M^+_1(\Omega)}

\newcommand{\Po}{\mathcal{P}}

\newcommand{\D}{\mathcal{D}}

\begin{document}

\title[Optimal measurements]{Optimal measurements in quantum mechanics}

\author{Teiko Heinonen}
\address{Teiko Heinonen, Department of Physics, University of Turku, 
FIN-20014 Turku, Finland}
\email{teiko.heinonen@utu.fi}

\begin{abstract}
Four common optimality criteria for measurements are formulated using
relations in the set of observables, and their connections are
clarified. As case studies, $1-0$ observables, localization
observables, and photon counting observables are considered. 
\\

\end{abstract}

\maketitle

\section{Introduction}

Any measurement is carried out in order to gain information about
an object system. Informationally complete measurements
\cite{Prugovecki77} allow a unique determination of the state of the
object, and therefore, they are usually regarded as optimal
measurements. Informationally complete phase space measurements are
well known \cite{AP77a} (also see \cite{OQP,QMPS,CRQM}), and other instances of
informationally complete measurements have been found as well; see,
for instance, \cite{RBSC04, DPS04}. However, in many practical cases a
unique state determination is not attainable. For example, a
photodetection or a position measurement does not provide enough
information for that purpose. It is still meaningful to
seek an optimal measurement in these cases, i.e., a measurement that
gives as much information as possible. The optimality of a measurement
depends on a specified class of measurements under investigation, and
it is therefore a relative property. The specified class of
measurements is determined by the requirements and presumptions
concerning measurements. Measurements may be, for example, required to
be covariant with respect to a relevant symmetry group.    

In addition to providing as much information as possible, it would be
desirable for a measurement to have as little imprecision as possible. This
objective can be thought just as another criterion for an optimal
measurement, and it has been investigated in \cite{MdM90a,DdG97,BDKPW05}. 

An imprecise measurement cannot give more information than
a more precise counterpart. However, in some cases it may be equally
good in state determination or in state distinction. This simple fact is
important since imprecision is unavoidable in any real measurement.
     
In this paper we study measurements only in the aspect of
measurement outcome statistics, and therefore, for our purposes
a measurement may be described by an observable (normalized positive
operator measure). We emphasize that this is only a partial
description of a measurement as, for instance, a possible preparative
purpose of measurements is ignored. Obviously, consideration of the
other aspects of measurements would give different optimality criteria.

The concept of an observable is briefly reviewed in
Section~\ref{Observables}, where we also recall the description of an
observable as an affine mapping from the set of states into the set of
probability measures. In Sections~\ref{Relations} and \ref{Optimal} we
formulate four common optimality criteria using certain relations on the set of
observables. Two of these relations correspond to the state distinction
and determination, while the other two are related to the measurement 
imprecision. These relations are pre-orderings, and thus,
they define partial orderings in the respective sets of equivalence
classes. The optimality criteria are then defined as maximality
requirements for equivalence classes. This approach is suitable also
for cases where informationally complete observables does not exist,
and connections between different criteria are easily seen.     
In Section~\ref{Examples} we study the cases of $1-0$ observables,
photon counting observables, and localization observables.

\section{Observables in quantum mechanics}\label{Observables}   

In this section we fix the notation, and for the reader's convenience
we briefly recall the basic description of a quantum observable. 
(For a review see, for example, \cite{OQP,PSAQT,QTCM}).

Let $\hi$ be a complex separable Hilbert space, and denote
the set of bounded linear operators on $\hi$ by $\lh$.  
Let $\Omega$ be a set and $\A$ a $\sigma$-algebra on
$\Omega$. The set of probability measures on the measurable space
$(\Omega,\A)$ is denoted by $\tn$. 

Consider a quantum system, described by a Hilbert space $\hi$.
\emph{States} of the system are represented as positive operators of
trace one, and \emph{observables} are represented as normalized
positive operator measures. More precisely, an observable with an outcome space
$(\Omega,\A)$ is a mapping $E:\A\to\lh$ such that  
\begin{itemize}
\item[(i)] $E(X)\geq O$ for any $X\in\A$;
\item[(ii)] $E(\Omega)=I$;
\item[(iii)] $E(\cup_i X_i)=\sum_i E(X_i)$
  (in the weak sense) for any disjoint sequence $(X_i)\subset\A$. 
\end{itemize}
We denote the set of states by $\sh$ and the set of observables
with the outcome space $(\Omega,\A)$ by $\obsoa$, or just $\obso$ when
$\A$ is understood.

Let $E\in\obsoa$ be an observable. For a state $T\in\sh$,
we define a probability measure $p^E_T$ on $(\Omega,\A)$ by
\begin{equation*}
p^{E}_T(X)=\tr [TE(X)],\quad X\in\A.
\end{equation*}
This is interpreted as the probability distribution of measurement
outcomes when the system is in the state $T$ and the observable $E$ is
measured. The observable $E$ defines a mapping $\Phi_E$ from $\sh$ to
$\tn$ by $\Phi_E(T)=p^E_T$. The mapping $\Phi_E$ preserves convex
combinations of states: for any $T_1,T_2\in\sh$ and $0\leq\lambda\leq
1$, we have 
\begin{equation}\label{convex}
\Phi_E(\lambda T_1 + (1-\lambda) T_2)= \lambda \Phi_E(T_1) +
(1-\lambda)\Phi_E(T_2).
\end{equation} 
Conversely, a mapping $\Phi:\sh\to\tn$ satisfying
(\ref{convex}) defines a unique observable $E_{\Phi}$. This
correspondence is consistent in the sense that $E_{\Phi_E}=E$ and
$\Phi_{E_{\Phi}}=\Phi$. For reviews of the properties of the mapping
$\Phi_E$, we refer to \cite{CL93,BCL95,BB97}.

The representation of an observable via an affine mapping from the set
of states $\sh$ into the space of probability measures $\tn$ is
physically natural. It captures an intuitive concept of an observable:
a specification of the outcome space (possible events in a  
measurement) and an assignment of a probability distribution to each
state of the system. In the following sections we use this
representation of observables to make the operational content of
the relations and the optimality criteria transparent.

\section{Relations on the set $\obsoa$}\label{Relations}

\subsection{State distinction and state determination}\label{State}

Let us first recall the usual concepts related to the ability of an
observable to distinguish and determine states. (For more details, see
e.g. \cite{BL89}.)  

\begin{definition}
Let $E\in\obsoa$ and $T_1,T_2\in\sh$.
\begin{itemize}
\item[(i)]
$E$ \emph{distinguish} the states $T_1$ and $T_2$ if 
\begin{equation*}
\Phi_{E}(T_1)\neq \Phi_{E}(T_2);
\end{equation*}
\item[(ii)]
the state $T_1$ is \emph{determined} by $E$ if, for all $T\in\sh$,
\begin{equation*}
\Phi_{E}(T_1)=\Phi_{E}(T)\ \Rightarrow T_1=T.
\end{equation*}
\end{itemize}
We denote by $\D_E$ the set of states determined by $E$. 
\end{definition}

The first of these concepts leads to the following relations
\cite{Davies70}. 

\begin{definition}\label{sdp}
Let $E,F\in\obsoa$. If for all states $T_1,T_2\in\sh$,
\begin{equation}\label{state1}
\Phi_{E}(T_1)=\Phi_{E}(T_2)\ \Rightarrow \Phi_{F}(T_1)=\Phi_{F}(T_2),
\end{equation}
then we denote $F\info E$, and say that the \emph{state distinction power}
 of $E$ is greater than or equal to $F$ (or that $F$ gives less or the
 same information than $E$). If $F \info E \info F$, we say that
$E$ and $F$ are \emph{informationally equivalent}, and denote $E\isim F$.  
\end{definition}

Condition~(\ref{state1}) can be written in an equivalent form 
\begin{equation*}
\Phi_{F}(T_1)\neq \Phi_{F}(T_2)\ \Rightarrow \Phi_{E}(T_1)\neq\Phi_{E}(T_2).
\end{equation*}
Hence, $F\info E$ means that $E$ distinguish all states that are
distinguished by $F$. It is clear that $\info$ is a reflexive and
transitive relation, and therefore, $\isim$ is an equivalence relation. 

\begin{definition}\label{deter}
Let $E,F\in\obsoa$. If $\D_F\subseteq\D_E$, then we denote $F\de E$,
and say that the \emph{state determination power} of $E$ is greater
than or equal to $F$.   
\end{definition}

It is immediately seen that the relation $\de$ is reflexive and
transitive, and thus, it defines an equivalence relation $\dsim$ in
the natural way. 

We note that if $F\info E$ holds, then $F\de E$. Indeed, let
$T_1\in\D_F$, and let $T$ be a state such that
$\Phi_E(T)=\Phi_E(T_1)$. Relation $F\info E$ implies that
$\Phi_F(T)=\Phi_F(T_1)$, and thus, $T_1=T$. This means that
$T_1\in\D_E$, and therefore $\D_F\subseteq\D_E$.    

Examples~\ref{trivial1} and \ref{sharpdet} show that the converse is,
in general, not true: the condition $F\de E$ does not imply that $F\info E$. 

\begin{example}\label{trivial1}
An observable $E\in\obsoa$ is \emph{trivial} (or \emph{uninformative})
if it does not distinguish any pair of states, that is,  
\begin{equation}\label{tri}
\Phi_E(T_1)=\Phi_E(T_2)\quad \forall T_1,T_2\in\sh.
\end{equation}
Condition~(\ref{tri}) is equivalent with the fact that there is a
probability measure $m\in\tn$ such that $E(X)=m(X)I$. If $E$ is a
trivial observable, then obviously $E\info F$ for any
$F\in\obsoa$. Moreover, if $F\info E$, then also $F$ is a trivial observable.  
\end{example}

\begin{example}\label{sharpdet}
Suppose that $F\in\obsoa$ is a spectral measure, i.e., $F(X)^2=F(X)$ for any
$X\in\A$. It is shown in~\cite{BL89} that $T\in \D_F$ if and
only if $T$ is a one-dimensional spectral projection of $F$, that is,
$T=F(X)=\ketbra{\psi}{\psi}$ for some unit vector $\psi\in\hi$. Thus,
if $F$ has no non-degenerate eigenstates, then $\D_F=\emptyset$. For
any trivial observable $E$ we also have $\D_E=\emptyset$, and hence,
$F\dsim E$.   
\end{example}

\subsection{Fuzzy observables and coarse-graining}\label{Fuzzy}

Fuzzy sets are used in many different situations to
model imprecision and uncertainty, and they are also applicable to
describe imprecision in a measurement. We recall that a \emph{fuzzy
  set} in $\Omega$ is a function $\widetilde{X}$ from $\Omega$ to the
interval $[0,1]$, and the value $\widetilde{X}(\omega)$ represents the
degree of membership of $\omega$ in $\widetilde{X}$
\cite{Zadeh65,Zadeh68}. We identify a subset $X$ of $\Omega$ with the
characteristic function $\chi_X$, and in this way the subsets of
$\Omega$ are (special types of) fuzzy sets. A fuzzy set is called a
\emph{fuzzy event} if it is measurable, and we denote by  
$\widetilde{\A}$ the collection of fuzzy events. If $m\in\tn$ and
$\widetilde{X}\in\widetilde{\A}$, then the probability
$m(\widetilde{X})$ is defined by the integral  
\begin{equation}\label{fint}
m(\widetilde{X})=\int \widetilde{X}(\omega)\ dm(\omega).
\end{equation}

Measurement imprecision may be modelled by a mapping $\Lambda$ from
$\A$ to $\widetilde{\A}$. We require that 
\begin{itemize}
\item[(i)] $\Lambda(X')=\chi_{\Omega}-\Lambda(X)$;
\item[(ii)] $\sum_{i=1}^{\infty}\Lambda(X_i)=\chi_{\Omega}$ if
 $\cup_{i=1}^{\infty} X_i=\Omega$ and $X_i\cap X_j=\emptyset$ for all
 $i\neq j$.
\end{itemize} 
Condition (i) means that a complement of a set is mapped to a fuzzy
complement, while (ii) means that a partition of $\Omega$ is mapped to a fuzzy
partition. We call a mapping $\Lambda:\A\to\widetilde{\A}$ with
properties (i) and (ii) a \emph{confidence mapping}.  

Suppose that $\Lambda$ is a confidence mapping and let $m\in\tn$. In
view of (\ref{fint}), the composite mapping $m\circ
\Lambda$ makes sense. The properties (i) and (ii) of $\Lambda$ imply
that $m\circ \Lambda$ is a probability measure. Our consideration
leads to the following definition. 

\begin{definition}
Let $E,F\in\obsoa$. If there exists a confidence mapping
$\Lambda:\A\to\widetilde{\A}$ such that, for any $T\in\sh$, 
\begin{equation}\label{fuzzy1}
\Phi_F(T)=\Phi_E(T)\circ \Lambda,
\end{equation}
then we denote $F\fuz E$ and say that $F$ is \emph{fuzzy version} of $E$.  
If $F\fuz E\fuz F$, we denote $F\fsim E$.
\end{definition}

There is an equivalent formulation of the relation $\fuz$. A mapping
\begin{equation*}
\nu:\Omega\times\A\to [0,1]
\end{equation*}
is a \emph{Markov kernel} if
\begin{itemize}
\item[(i)] for every $\omega\in\Omega$, the mapping
  $\nu(\omega,\cdot)$ is a probability measure on $(\Omega,\A)$;
\item[(ii)] for every $X\in\A$, the mapping $\nu(\cdot,X)$ is $\A$-measurable. 
\end{itemize}
It is straightforward to verify that $\nu$ is Markov kernel if and
only if the mapping $X\mapsto\nu(\cdot,X)$ is a confidence
mapping. Hence, the condition $F\fuz E$ is equivalent to the fact that
there exists a Markov kernel $\nu$ such that  
\begin{equation}\label{fuzzy2}
F(X)=\int\nu(\omega,X)\ dE(\omega),\quad X\in\A.
\end{equation}
   
A formulation similar to (\ref{fuzzy2}) was introduced in
\cite{AE74,AD76,AP77b}, and it has been used, for
instance, to investigate joint position-momentum measurements. The
relation $\fuz$ has been studied in~\cite{MdM90a} in the case of finite
dimensional Hilbert spaces and countable outcome spaces.
The general case (with a slightly different relation than ours) has
been studied in~\cite{DdG97}.     

The relation $\fuz$ is reflexive since the mapping 
\begin{equation*}
(\omega,X)\mapsto \delta_{\omega}(X)=\chi_X(\omega)
\end{equation*}
is a Markov kernel and
\begin{equation*}
E(X)=\int \chi_X(\omega)\ dE(\omega).
\end{equation*}

\begin{proposition}
The relation $\fuz$ is transitive.
\end{proposition}

\begin{proof}
Let $F_i\in\obsoa$, $i=1,2,3,$ and assume that $F_1\fuz F_2$ and
$F_2\fuz F_3$, with $\nu_1$ and $\nu_2$ being corresponding Markov
kernels, respectively. For any $\omega\in\Omega,X\in\A$, define
$$
\nu_3(\omega,X)=\int\nu_1(\omega',X)\ \nu_2(\omega,d\omega').
$$
Let us first note that $\nu_3$ is a Markov kernel. Indeed, for a fixed
$X\in\A$, the function $\nu_1(\cdot,X)$ is nonnegative, bounded and
measurable. Therefore, there is an increasing sequence $\{h_n\}$ of
nonnegative simple functions converging to the function $\nu_1(\cdot,X)$
pointwisely. For each $\omega\in\Omega$, the monotone convergence
theorem implies that
\begin{equation*}
\int\nu_1(\omega',X)\ \nu_2(\omega,d\omega')=\lim_{n\to\infty}
\int h_n(\omega')\ \nu_2(\omega,d\omega').
\end{equation*}  
For every $n$, the function $\omega\mapsto \int h_n(\omega')\
\nu_2(\omega,d\omega')$ is measurable and the function
$\nu_3(\cdot,X)$ is a pointwise limit of measurable
functions. Hence, the function $\nu_3(\cdot,X)$ is measurable.  It is easy
to see that, for a fixed $\omega\in\Omega$, the mapping
$\nu_3(\omega,\cdot)$ is a probability measure. In conclusion, $\nu_3$ is a
Markov kernel.  

Let $T\in\sh$. For any $X\in\A$, we have
\begin{eqnarray*}
\int \nu_3(\omega,X)\ dp^{F_3}_T(\omega) &=&
\int\int\nu_1(\omega',X)\ \nu_2(\omega,d\omega')
\ dp^{F_3}_T(\omega)\\
&=& \lim_{n\to\infty}
\int\int h_n(\omega')\ \nu_2(\omega,d\omega')\ dp^{F_3}_T(\omega)\\
&=& \lim_{n\to\infty}
\int h_n(\omega')\ dp^{F_2}_T(\omega) = \int \nu_1(\omega',X)\
dp^{F_2}_T(\omega')\\ 
&=& p^{F_1}_T(X).
\end{eqnarray*}
This shows that $F_1\fuz F_3$.
\end{proof}

\begin{example}\label{trivial2}
Let $E\in\obsoa$ be a trivial observable defined by a
probability measure $m\in\tn$; see Example~\ref{trivial1}. For any
$F\in\obsoa$, we then have $E\fuz F$. Indeed, define
$$
\nu(\omega,X)=m(X),\quad \omega\in\Omega, X\in\A.
$$
Then $\nu$ is a Markov kernel and
$$
\int \nu(\omega,X)\ dF(\omega) = m(X)\int dF(\omega)=m(X)I=E(X).
$$
Moreover, it is easy to see that if $F\fuz E$, then
also $F$ is a trivial observable.
\end{example}

Suppose that $F\fuz E$, and let $\nu$ be a corresponding Markov
kernel such that~(\ref{fuzzy2}) holds. Define a mapping
$\Psi_{\nu}:\tn\to\tn$ by \begin{equation}\label{Psinu}
\Psi_{\nu}(m)(X)=\int \nu(\omega,X)\ d m(\omega), \quad
m\in\tn, X\in\A.
\end{equation}
From equation~(\ref{fuzzy2}) follows that $\Phi_F$ is a composite
mapping of $\Phi_E$ and $\Psi_{\nu}$, that is,
\begin{equation*}
\Phi_F=\Psi_{\nu}\circ\Phi_E.
\end{equation*} 
Hence, any measurement outcome distribution of the
observable $F$ is obtained from the corresponding measurement outcome
distribution of $E$ by applying a mapping $\Psi_{\nu}$, which is
independent of a state. This procedure is formulated in the following
concept of coarse-graining. The concept of coarse-graining means,
generally speaking, a reduction in the statistical description of a
system; see, for instance, \cite{BQ93}. 

\begin{definition}
Let $E,F \in\obsoa$. We say that $F$ is a \emph{coarse-graining} of
$E$, and denote $F\coar E$, if there exists an affine mapping
$\Psi:\tn \to \tn$ such that  
\begin{equation}\label{fpsie}
\Phi_F=\Psi \circ \Phi_E.
\end{equation}
\end{definition}

The relation $\coar$ is reflexive as the identity mapping is
affine, and the transitivity of $\coar$ follows from the fact that the
composition of affine mappings is affine. The corresponding
equivalence relation is denoted by $\csim$.

Our previous discussion shows that if $F\fuz E$, then $F\coar E$. We
note that there are affine mappings on $\tn$ which do not have
representations via Markov kernels as in (\ref{Psinu}); see \cite{BHS98}. 
However, for observables on a finite outcome space the relations
$\fuz$ and $\coar$ are the same, as the following example illustrates. 

\begin{example}\label{finite}
Suppose that $\Omega=\{1,2,\ldots,n\}$. An observable $E\in\obso$ is
determined by the effects $E_j:=E(\{j\})$, and for each Markov kernel
$\nu$ corresponds a $n\times n$ row stochastic matrix $(\nu_{jk})$, where
$\nu_{jk}=\nu(j,\{k\})$. Condition (\ref{fuzzy2}) can then be written
in the form 
\begin{equation}\label{fuzzyn}
F_k=\sum_{j=1}^n \nu_{jk} E_j,\quad k\in\Omega.
\end{equation}
For an affine mapping $\Psi$
on $\tno$, define $\nu(j,X):=\Psi(\delta_j)(X)$, where $\delta_j$ is the point
measure concentrated at a point $j\in\Omega$ and $X\subseteq\Omega$. Since any probability measure on $\Omega$ can be written as a
convex combination of the point measures, the mapping $\Psi$ is
determined by the Markov kernel $\nu$. We conclude that $F\coar E$ if
and only if $F\fuz E$, and this is the case exactly when there is a
stochastic matrix such that (\ref{fuzzyn}) holds. 
\end{example}

The condition $F\coar E$ implies that $F\info E$. Indeed, if there is
a mapping $\Psi$ such that (\ref{fpsie}) holds, then certainly
condition~(\ref{state1}) is satisfied.

\section{Optimal measurements}\label{Optimal}

Let $\preccurlyeq$ be one of the relations $\fuz,\coar,\info$ or $\de$, and let
$\sim$ be the corresponding equivalence relation. Since $\preccurlyeq$
is reflexive and transitive, it defines a partial ordering
$\preccurlyeq'$ on the set of equivalence classes $\obsoa/\sim$.
Namely, denoting the equivalence class of an observable $E$ by $[E]$,
we define  
\begin{equation*}
[E]\preccurlyeq' [F]\ \textrm{if and only if}\ E\preccurlyeq F.
\end{equation*}
Typically, we have some requirements and presumptions for the
intended measurements, and therefore, we are interested only on a
restricted class $\ob\subseteq\obsoa$ of observables. We are thus led
to the following definition.

\begin{definition}\label{opti}
Let $\ob\subseteq\obsoa$. We say that an observable $E\in\ob$ is \emph{optimal} in $\ob$ with respect to preordering $\preccurlyeq$ (or $\preccurlyeq$-\emph{optimal} in $\ob$), if the equivalence class of $E$ is a maximal element of the partially ordered set $\ob/\sim$. 
\end{definition}

In other words, $E$ is $\preccurlyeq$-optimal in $\ob$ if, for
any $F\in\ob$, the condition $E\preccurlyeq F$ implies that $E\sim F$. 

It was shown in the last section that, for
observables $E$ and $F$, the following implications hold: 
\begin{equation}\label{implic1}
F\fuz E\ \Rightarrow\  F\coar E\ \Rightarrow\
F\info E\ \Rightarrow F\de E.
\end{equation}
This means also that the following inclusions hold between the
equivalence classes of $E$: 
\begin{equation}\label{inclus1}
[E]_f\subseteq [E]_c\subseteq [E]_i\subseteq [E]_d.
\end{equation}

We emphasize that although the relations have the hierarchy (\ref{implic1}), a $\fuz$-optimal observable may or may not be optimal with respect to other relations. This is demonstrated in
Section~\ref{Examples}. However, if an observable $E\in\ob$ satisfies a
stronger condition that $F\fuz E$ for any $F\in\ob$ (i.e., the
equivalence class $[E]$ is the greatest element), then it follows that $E$ is
optimal in $\ob$ with respect to all four relations.

We note that the four relations discussed here are not the only interesting relations in the theory of quantum measurements. In the recent paper \cite{BDKPW05} several other relations were studied, and the notion of  a \emph{clean} measurement was defined similarly to Definition \ref{opti}. Cleanness property is also a relevant optimality criterion.

\section{Examples}\label{Examples}

\subsection{1-0 observables}\label{Simple}

The set of \emph{effects}, denoted by $\eh$, is the set of operators
$A\in\lh$ satisfying $O\leq A\leq I$. An effect $A$ defines an
observable $E^A$ with the outcome space $\Omega=\{0,1\}$ by
\begin{equation*}
E^A_1=A,\ E^A_0=A'\equiv I-A.
\end{equation*}
These are the most simplest kind of observables, and we call them
\emph{1-0 observables}. 

\begin{proposition}\label{ynorder}
Let $A,B\in\eh$ and let $E^A$, $E^B$, be the corresponding 1-0 observables. 
Then $E^A\fuz E^B$ if and only if there are numbers $s,t\in [0,1]$ such that
\begin{equation}\label{st}
A=tB+sB'.
\end{equation}
\end{proposition}

\begin{proof}
As shown in Example~\ref{finite}, the condition $E^A\fuz E^B$ means
that there is a row stochastic matrix $(\nu_{jk})$ such that
\begin{eqnarray*}
E^A_0 &=& \nu_{00}\ E^B_0 + \nu_{10}\ E^B_1,\\
E^A_1 &=& \nu_{01}\ E^B_0 + \nu_{11}\ E^B_1.
\end{eqnarray*}
Since $\nu_{11}+\nu_{10}=\nu_{01}+\nu_{00}=1$, these equations are
equivalent. Therefore, the condition $E^A\fuz E^B$ holds if and only if
\begin{equation*}
A=\nu_{11}\ B+\nu_{01}\ B'.
\end{equation*}
Any $2\times 2$ row stochastic matrix has the form
\begin{equation*}
 \nu_{11}=t,\quad \nu_{10}=1-t,\quad \nu_{01}=s,\quad \nu_{00}=1-s, 
\end{equation*}
for some numbers $s,t\in [0,1]$, and thus, the claim follows.
\end{proof}

As a direct consequence of Proposition~\ref{ynorder}, we note that, 
for non-trivial observables $E^A$ and $E^B$, the equivalence relation
$E^A\fsim E^B$ holds exactly when $A=B$ or $A=B'$.   

\begin{proposition}\label{ynopt}
Let $A\in\eh$. The observable $E^A$ is $\fuz$-optimal in $\obso$ if
and only if $\no{A}=\no{A'}=1$.  
\end{proposition} 

\begin{proof}
Let us first assume that $\no{A}=\no{A'}=1$. Suppose that $B$ is an effect such
that $E^A\fuz E^B$. We need to show that $E^A\fsim E^B$. By
Proposition~\ref{ynorder} the condition $E^A\fuz E^B$ is equivalent
with the fact that there exist numbers $s,t\in[0,1]$ such that
(\ref{st}) holds. Since $\no{A'}=1$, for any $\epsilon >0$ there is
a unit vector $\varphi_{\epsilon}\in\hi$ such that 
\begin{equation*}
\ip{\varphi_{\epsilon}}{(I-A)\varphi_{\epsilon}}\geq 1-\epsilon,
\end{equation*}
and thus, 
\begin{equation}\label{Aeps}
\ip{\varphi_{\epsilon}}{A\varphi_{\epsilon}}\leq\epsilon.
\end{equation}
From (\ref{st}) and (\ref{Aeps}) we get
\begin{eqnarray*}
\epsilon & \geq &
s(1-\ip{\varphi_{\epsilon}}{B\varphi_{\epsilon}}) + 
t\ip{\varphi_{\epsilon}}{B\varphi_{\epsilon}}\\
& \geq & \min (s,t).
\end{eqnarray*}
Thus, either $s=0$ or $t=0$. If $s=0$, then $A=tB$. Moreover, as
\begin{equation*}
1=\no{A}=t\no{B}\leq t\leq 1,
\end{equation*}
we have $t=1$ and $A=B$. By a similar argument $t=0$ gives $A=B'$.  
Thus, $E^A\fsim E^B$.
 
Let us then assume that $\no{A}<1$ (the case $\no{A'}<1$ being
similar). Denote $\alpha:=\no{A}$ and $\beta:=\no{A'}$. Then 
\begin{equation}\label{bAa}
(1-\beta)I\leq A \leq \alpha I
\end{equation}
and
\begin{equation*}
\alpha+\beta=\no{A}+\no{A'}\geq \no{A+A'}=1.
\end{equation*}
If $\alpha+\beta=1$, then (\ref{bAa}) implies that  $A=\alpha I$. In
this case $E^A$ is a trivial observable, and clearly, not $\fuz$-optimal.
Consider the case $\alpha+\beta>1$. It follows from (\ref{bAa}) that
the operator 
$$
B:=\frac{1}{\alpha+\beta-1} A + \frac{\beta-1}{\alpha+\beta-1}I
$$
is an effect. Moreover,
$$
A=tB+sB',
$$
where $s=1-\beta$ and $t=\alpha$. Thus, $E^A\fuz E^B$.  Since
$0<\alpha<1$, we have $B\neq A\neq B'$. This shows that $E^A$ is not
$\fuz$-optimal.
\end{proof}

The set $\obso$ is convex: if $E^A,E^B\in\obso$ and
$0\leq\lambda\leq 1$, then  
\begin{equation*}
\lambda E^A+(1-\lambda)E^B = E^{\lambda A+(1-\lambda)B}\in\obso.
\end{equation*}
If $A\neq B$ and $0<\lambda<1$, then the convex combination
$E^{\lambda A+(1-\lambda)B}$ is a \emph{randomized observable}
\cite{PSAQT}. An observable is \emph{non-randomized} if it has no
such convex decomposition. The
extreme elements of the convex set $\eh$ are projection operators
\cite[Lemma 2.3]{QTOS},
and hence, an observable $E^A$ is non-randomized exactly when the
respective effect $A$ is a projection. That kind of
observables are $\fuz$-optimal in $\obso$, but if $\dim(\hi)\geq 3$,
then there are also other $\fuz$-optimal observables. To give an
example, let $P$ and $R$ be mutually orthogonal one-dimensional
projections. For any $0<t<1$, the operator $A=P+t R$ is an effect but
not a projection, and $\no{A}=\no{A'}=1$. The observable
$E^A$ is a convex combination of the non-randomized observables $E^P$ and
$E^{P+R}$, and all these three observables are $\fuz$-optimal.

\begin{remark}
The condition $\no{A}=\no{A'}=1$ in Proposition~\ref{ynopt} has a
physical interpretation. Indeed, if $P$ is a projection (and not equal
to $O$ or $I$), then there exist states $T_1$ and $T_2$ such that 
\begin{equation}\label{real}
\tr [T_1P]=1, \quad \tr [T_2P']=1.
\end{equation}
This means that $P$ and $P'$ can be realized in the states $T_1$ and
$T_2$, and thus, they are \emph{actualizable properties}.
On the other hand, the condition $\no{A}=\no{A'}=1$ is equivalent with the
fact that for each $\delta >0$ there exist states $T_1$ and $T_2$ such that
\begin{equation}\label{appreal}
\tr [T_1A]\geq 1-\delta,\quad \tr [T_2A']\geq 1-\delta.
\end{equation}
This is a relaxation of (\ref{real}), and we say that the
effects $A$ and $A'$ are \emph{approximately actualizable properties}. 
\end{remark}

\subsection{Photon counting observables}\label{Photon}

Let $\hi$ be a Hilbert space describing a one-mode of an
electromagnetic field. We denote by $\N$ the set of natural numbers
(including $0$), and $\Po(\N)$ is the set of all subsets of $\N$. 
Given an observable $F$ with the outcome space $(\N,\Po(\N))$, we
denote $F_n=F(\{n\})$. Also, if $\nu:\N\times\Po(\N)\to[0,1]$ is a
Markov kernel, we denote $\nu_{kn}=\nu(k,\{n\})$, $k,n\in\N$.  

The number operator $N=a^*a$ has a non-degenerate eigenvector
$\ket{n}$ for every $n\in\N$. The number observable $E^N$ with the
outcome space $(\N,\Po(\N))$ is defined by   
\begin{equation*}
E^N_n=\ketbra{n}{n},\quad n\in\N.
\end{equation*}
A photodetector with efficiency $\epsilon$, $0\leq\epsilon\leq 1$, may be
  described by an observable $F^{\epsilon}$ defined by 
\begin{equation}\label{pco}
F^{\epsilon}_n=\sum_{m=n}^{\infty} {m\choose n}\epsilon^n (1-\epsilon)^{m-n}
\ketbra{m}{m},\quad n\in\N,
\end{equation}
see, e.g., \cite[Section VII.3.]{OQP}. We denote by $\ob_{\mathcal{P}}$
the set of this kind of observables, and we call them \emph{photon
counting observables}. The photon counting observable $F^1$
corresponding to the ideal efficiency $\epsilon=1$ is the number
observable $E^N$, and the observable $F^0$ is the trivial observable
given by $F^0_n=\delta_{0,n} I$. 

In the following we investigate the set $\ob_{\mathcal{P}}$ of photon counting observables. Some related results have been discussed in \cite[Chapter 7]{FQMEA}.

\begin{proposition}\label{eporder}
Let $F^{\epsilon_1},F^{\epsilon_2}\in\ob_{\mathcal{P}}$. The condition 
$F^{\epsilon_1}\fuz F^{\epsilon_2}$ holds if and only if
$\epsilon_1\leq\epsilon_2$.
\end{proposition}

\begin{proof}
Let us first assume that $F^{\epsilon_1}\fuz F^{\epsilon_2}$. This
means that there exists a Markov kernel $\nu$ such that  
\begin{equation*}
F^{\epsilon_1}_n=\sum_{k=0}^{\infty}\nu_{kn} F^{\epsilon_2}_k,\quad n\in\N.
\end{equation*}
For every $m,n\in\N$, we get 
\begin{equation}\label{eporder1}
\bra{m}F^{\epsilon_1}_n\ket{m}=\sum_{k=0}^{\infty}\nu_{kn}
\bra{m}F^{\epsilon_2}_k\ket{m}.
\end{equation}
Substituting (\ref{pco}) into both sides of (\ref{eporder1}) shows that
$\nu_{mm}=\epsilon_1^m \epsilon_2^{-m}$. Since $\nu$
is a Markov kernel, we have $\nu_{mm}\leq 1$. This can hold only if
$\epsilon_1\leq \epsilon_2$. 

Let us then assume that $\epsilon_1\leq\epsilon_2$. Define 
\begin{equation*}
\nu_{kn}=\left\{ \begin{array}{ll}
0 & \textrm{if } k<n, \\
{k\choose n}\epsilon_1^n\epsilon_2^{-k} (\epsilon_2-\epsilon_1)^{k-n}
& \textrm{if } k\geq n. 
\end{array}\right.
\end{equation*}
Then $\nu$ is a Markov kernel, and we have
\begin{eqnarray*}
\sum_{k=0}^{\infty} \nu_{kn} F^{\epsilon_2}_k &=&
\sum_{k=n}^{\infty} \sum_{m=k}^{\infty} {k\choose n}{m\choose k}
\epsilon_1^n (\epsilon_2-\epsilon_1)^{k-n}(1-\epsilon_2)^{m-k}
\ketbra{m}{m} \\
&=& \sum_{m=n}^{\infty} \epsilon_1^n \left( \sum_{k=n}^m {k\choose n}{m\choose
    k}(\epsilon_2-\epsilon_1)^{k-n} (1-\epsilon_2)^{m-k} \right)
\ketbra{m}{m} \\
&=& \sum_{m=n}^{\infty} {m\choose n} \epsilon_1^n
(1-\epsilon_1)^{m-n}\ketbra{m}{m} = F^{\epsilon_1}_n.
\end{eqnarray*}
Thus, $F^{\epsilon_1}\fuz F^{\epsilon_2}$.
\end{proof}

\begin{corollary}
The number observable $E^N$ is an optimal observable in
$\ob_{\mathcal{P}}$ with respect to $\fuz,\coar,\info$ and $\de$. 
\end{corollary}

Next we show that imprecision in a photon counting
measurement does not imply a loss of information.  

\begin{proposition}\label{Ni}
If $F^{\epsilon}\in\ob_{\mathcal{P}}$ and $\epsilon\neq 0$, then
$F^{\epsilon}\isim E^N$. 
\end{proposition}

\begin{proof}
As the claim is trivial in the case $\epsilon=1$, we may assume that
$0<\epsilon<1$. Moreover, since $F^{\epsilon}\fuz E^N$ by
Proposition~\ref{eporder}, we have $F^{\epsilon}\info E^N$. To prove
that $E^N\info 
F^{\epsilon}$, let $T_1,T_2\in\sh$ and assume that
$\Phi_{F^{\epsilon}}(T_1)=\Phi_{F^{\epsilon}}(T_2)$. By (\ref{pco})
this means that, for every $n\in\N$,
\begin{equation}\label{Ni1}
\sum_{m=n}^{\infty} {m\choose n} (1-\epsilon)^{m}
\bra{m}T_1-T_2\ket{m}=0.
\end{equation}
Denote $a_m:= (1-\epsilon)^m \bra{m}T_1-T_2\ket{m}$ for every $m\in\N$.
Since $|a_m|\leq (1-\epsilon)^m$, the formula 
\begin{equation*}
f(z):= \sum_{m=0}^{\infty} a_m z^m
\end{equation*}
defines a holomorphic function in the region
$|z|<\frac{1}{1-\epsilon}$. The $n$th derivative of $f$ is
\begin{equation*}
f^{(n)}(z)= \sum_{m=n}^{\infty}m(m-1)\cdots(m-n+1) a_m z^{m-n}, 
\end{equation*}
and hence, (\ref{Ni1}) implies that $f^{(n)}(1)=0$ for every
$n\in\N$. Thus, $f=0$, and $a_m=0$ for every $m\in\N$. We
conclude that $\Phi_{E^N}(T_1)=\Phi_{E^N}(T_2)$, and therefore,
$E^N\info F^{\epsilon}$.  
\end{proof}
       
\begin{corollary}
If $F^{\epsilon}\in\ob_{\mathcal{P}}$ and $\epsilon\neq 0$, then
$\D_{F^{\epsilon}}=\{\ketbra{n}{n}\mid n\in\N\}$. 
\end{corollary}

\begin{proof}
For the number observable $E^N$ the claim follows from~\cite{BL89},
(see Example~\ref{sharpdet}). Since
$F^{\epsilon}\isim E^N$ by Proposition~\ref{Ni}, we have
$F^{\epsilon}\dsim E^N$, and thus, $\D_{F^{\epsilon}}=\D_{E^N}$.
\end{proof}

\subsection{Localization observables on $\R$}

Let us consider a free particle in the real line $\R$. We
denote by $U$ and $V$ be the one-parameter unitary representations
related to the groups of space translations and velocity boosts, 
respectively. As shown, for instance, in Chapter III of \cite{PSAQT},
we may fix $\hi=L^2(\R)$ and take $U$ and $V$ act on $\varphi\in\hi$ as 
\begin{eqnarray*}
\left[U(q)\varphi\right](x) &=& \varphi(x-q), \\
\left[V(p)\varphi\right](x) &=& e^{ipx}\varphi(x).
\end{eqnarray*}
Let $Q$ be the selfadjoint operator such that $V(p)=e^{ipQ}$ for every
$p\in\R$. The spectral measure $E^Q$ corresponding to the operator
$Q$ is an observable with the outcome space $(\R,\br)$, where $\br$ is
the Borel $\sigma$-algebra of $\R$. For any $X\in\br$ and
$\varphi\in\hi$, we have the usual formula
\begin{equation*}
E^Q(X)\varphi=\chi_X\varphi,
\end{equation*}
where $\chi_X$ is the characteristic function of $X$.

The observable $E^Q$ has the property that, for any $q\in\R,X\in\br$,  
\begin{equation}\label{covQ}
U(q)E^Q(X)U(q)^* = E^Q(X+q).
\end{equation}
This covariance property justifies to associate the observable $E^Q$ with a
localization measurement of the particle. In general, an
observable $F$ with the outcome space $(\R,\br)$ is a
\emph{localization observable} if it has the covariance property
\begin{equation}\label{covF}
U(q)F(X)U(q)^* = F(X+q),\quad q\in\R,X\in\br.
\end{equation}
We denote by $\loc$ the set of localization observables. 

\begin{proposition}\label{loc}
Let $F\in\loc$. The following conditions are
equivalent:
\begin{itemize}
\item[(i)] $F\info E^Q$;
\item[(ii)] $F\coar E^Q$;
\item[(iii)] $F\fuz E^Q$;
\item[(iv)] for every $p\in\R,X\in\br$, 
\begin{equation}\label{invar}
V(p)F(X)V(p)^*=F(X);
\end{equation}
\item[(v)] there is a probability measure $\rho\in M^+_1(\R)$ such that  
\begin{equation}\label{rho}
\Phi_F(T)=\rho\ast\Phi_{E^Q}(T),\quad T\in\sh,
\end{equation}
where $\rho\ast\Phi_{E^Q}(T)$ is the convolution of the measures
$\rho$ and $\Phi_{E^Q}(T)$.
\end{itemize}
\end{proposition}

\begin{proof}
It is shown in \cite{HLY04} that conditions (ii), (iii), and (v) are
equivalent, and (iv) and (v) are equivalent by \cite{CHT04}. Since
(ii)$\Rightarrow$(i), it is enough to show that
(i)$\Rightarrow$(iv). 

Assume (i). Let $\psi_1\in\hi$ be a unit vector,
$p\in\R$, and denote $\psi_2=V(p)^*\psi_1$. Let $T_1$ and $T_2$ be
the states corresponding to the vectors $\psi_1$ and $\psi_2$, respectively. A
short calculation shows that $\Phi_{E^Q}(T_1)=\Phi_{E^Q}(T_2)$,
and therefore, by the assumption we have
$\Phi_F(T_1)=\Phi_F(T_2)$. This means that  
\begin{equation}\label{invE}
\ip{\psi_1}{F(X)\psi_1}=\ip{\psi_1}{V(p)F(X)V(p)^*\psi_1}
\end{equation}
for all $X\in\br$. As $\psi_1$ was an arbitrary unit vector, (iv) follows.
\end{proof}

The condition~(\ref{invar}) means that the localization observable $F$
is invariant under velocity boosts. In $F$ satisfy both (\ref{covF})
and (\ref{invar}), it is called a \emph{position observable},
\cite{OQP,CHT04}. It is clear from Proposition~\ref{loc} that $E^Q$ is
an optimal position observable. However, not all localization
observables are position observables. The localization observables
have been characterized in \cite{Holevo83, CDT04}, and it is known
that there are localization observables which do not have the
invariance property~(\ref{invar}). It follows that there are
localization observables which do not satisfy the relations (i), (ii)
and (iii).  
   
\begin{proposition}
The observable $E^Q$ is $\fuz$-optimal in $\loc$.
\end{proposition}

\begin{proof}
Let $F\in\loc$ and assume that $E^Q\fuz F$. By Remark 3 of \cite{DdG97},
we then have $\ran{E^Q}\subseteq\ran{F}$. Since a projection in the
range of $F$ commutes with the other effects in the range (see
e.g. \cite{LP01}), we get 
\begin{equation*}
F(X)E^Q(Y)=E^Q(Y)F(X)
\end{equation*}
for all $X,Y\in\br$.
Thus, by the functional calculus we get
\begin{equation*}
F(X)V(p)=V(p)F(X)
\end{equation*}
for all $p\in\R,X\in\br$. This and Proposition~\ref{loc} imply that
$F\fuz E^Q$.   
\end{proof}

To author's knowledge it is not known whether the observable $E^Q$
is $\coar$-optimal or $\info$-optimal in $\loc$. Also, whether the
condition $\D_F = \emptyset$ holds for every $F\in\loc$ appears to be an
open question.

\section*{Acknowledgment}

The author would like to thank Pekka Lahti for many discussions and
his comments on an earlier version of this paper.


\begin{thebibliography}{10}

\bibitem{Prugovecki77}
E.~Prugove{\v{c}}ki.
\newblock Information-theoretical aspects of quantum measurements.
\newblock {\em Int. J. Theor. Phys.}, 16:321--331, 1977.

\bibitem{AP77a}
S.T. Ali and E.~Prugove{\v{c}}ki.
\newblock Classical and quantum statistical mechanics in a common {L}iouville
  space.
\newblock {\em Phys. A}, 89(3):501--521, 1977.

\bibitem{OQP}
P.~Busch, M.~Grabowski, and P.J. Lahti.
\newblock {\em Operational quantum physics}.
\newblock Springer-Verlag, Berlin, 1997.

\bibitem{QMPS}
F.E. Schroeck.
\newblock {\em Quantum mechanics on phase space}.
\newblock Kluwer Academic Publishers Group, Dordrecht, 1996.

\bibitem{CRQM}
W.~Stulpe.
\newblock {\em Classical representations of quantum mechanics related to
  statistically complete observables}.
\newblock Wissenschaft und Technik Verlag, Berlin, 1997.

\bibitem{RBSC04}
J.M. Renes, R.~Blume-Kohout, A.J. Scott, and C.M. Caves.
\newblock Symmetric informationally complete quantum measurements.
\newblock {\em J. Math. Phys.}, 45(6):2171--2180, 2004.

\bibitem{DPS04}
G.M. D'Ariano, P.~Perinotti, and M.F. Sacchi.
\newblock Informationally complete measurements and group representation.
\newblock {\em J. Opt. B: Quantum Semiclass. Opt.}, 6:S487--S491, 2004.

\bibitem{MdM90a}
H.~Martens and W.M. de~Muynck.
\newblock Nonideal quantum measurements.
\newblock {\em Found. Phys.}, 20(3):255--281, 1990.

\bibitem{DdG97}
S.V. Dorofeev and J.~de~Graaf.
\newblock Some maximality results for effect-valued measures.
\newblock {\em Indag. Mathem., N.S.}, 8(3):349--369, 1997.

\bibitem{BDKPW05}
F.~Buscemi, G.M. D'ariano, M.~Keyl, P.~Perinotti, and R.~Werner.
\newblock Clean positive operator valued measures.
\newblock {\em Preprint}, quant-ph/0505095, 2005.

\bibitem{PSAQT}
A.S. Holevo.
\newblock {\em Probabilistic and statistical aspects of quantum theory}.
\newblock North-Holland Publishing Co., Amsterdam, 1982.

\bibitem{QTCM}
A.~Peres.
\newblock {\em Quantum theory: concepts and methods}.
\newblock Kluwer Academic Publishers, Dordrecht, 1993.

\bibitem{CL93}
G.~Cassinelli and P.J. Lahti.
\newblock Spectral properties of observables and convex mappings in quantum
  mechanics.
\newblock {\em J. Math. Phys.}, 34(12):5468--5475, 1993.

\bibitem{BCL95}
P.~Busch, G.~Cassinelli, and P.J. Lahti.
\newblock Probability structures for quantum state spaces.
\newblock {\em Rev. Math. Phys.}, 7(7):1105--1121, 1995.

\bibitem{BB97}
E.G. Beltrametti and S.~Bugajski.
\newblock Effect algebras and statistical physical theories.
\newblock {\em J. Math. Phys.}, 38(6):3020--3030, 1997.

\bibitem{BL89}
P.~Busch and P.~Lahti.
\newblock The determination of the past and the future of a physical system in
  quantum mechanics.
\newblock {\em Found. Phys.}, 19(6):633--678, 1989.

\bibitem{Davies70}
E.B. Davies.
\newblock On the repeated measurements of continuous observables in quantum
  mechanics.
\newblock {\em J. Functional Analysis}, 6:318--346, 1970.

\bibitem{Zadeh65}
L.A. Zadeh.
\newblock Fuzzy sets.
\newblock {\em Information and Control}, 8:338--353, 1965.

\bibitem{Zadeh68}
L.~A. Zadeh.
\newblock Probability measures of fuzzy events.
\newblock {\em J. Math. Anal. Appl.}, 23:421--427, 1968.

\bibitem{AE74}
S.T. Ali and G.G. Emch.
\newblock Fuzzy observables in quantum mechanics.
\newblock {\em J. Math. Phys.}, 15:176--182, 1974.

\bibitem{AD76}
S.T. Ali and H.D. Doebner.
\newblock On the equivalence of nonrelativistic quantum mechanics based upon
  sharp and fuzzy measurements.
\newblock {\em J. Math. Phys.}, 17(7):1105--1111, 1976.

\bibitem{AP77b}
S.T. Ali and E.~Prugove{\v{c}}ki.
\newblock Systems of imprimitivity and representations of quantum mechanics on
  fuzzy phase spaces.
\newblock {\em J. Math. Phys.}, 18(2):219--228, 1977.

\bibitem{BQ93}
P.~Busch and R.~Quadt.
\newblock Concepts of coarse graining in quantum mechanics.
\newblock {\em Int. J. Theor. Phys.}, 32(12):2261--2269, 1993.

\bibitem{BHS98}
S.~Bugajski, K.-E. Hellwig, and W.~Stulpe.
\newblock On fuzzy random variables and statistical maps.
\newblock {\em Rep. Math. Phys.}, 41(1):1--11, 1998.

\bibitem{QTOS}
E.B. Davies.
\newblock {\em Quantum theory of open systems}.
\newblock Academic Press, London, 1976.

\bibitem{FQMEA}
W.M. de~Muynck.
\newblock {\em Foundations of quantum mechanics, an empiricist approach}.
\newblock Kluwer Academic Publishers, Dordrecht, 2002.

\bibitem{HLY04}
T.~Heinonen, P.~Lahti, and K.~Ylinen.
\newblock Covariant fuzzy observables and coarse-graining.
\newblock {\em Rep. Math. Phys.}, 53(3):425--441, 2004.

\bibitem{CHT04}
C.~Carmeli, T.~Heinonen, and A.~Toigo.
\newblock Position and momentum observables on {$\Bbb R$} and on {${\Bbb R}\sp
  3$}.
\newblock {\em J. Math. Phys.}, 45(6):2526--2539, 2004.

\bibitem{Holevo83}
A.S. Holevo.
\newblock Generalized imprimitivity systems for abelian groups.
\newblock {\em Sov. Math. (Iz. VUZ)}, 27:53--80, 1983.

\bibitem{CDT04}
G.~Cassinelli, E.~De~Vito, and A.~Toigo.
\newblock Positive operator valued measures covariant with respect to an
  abelian group.
\newblock {\em J. Math. Phys.}, 45(1):418--433, 2004.

\bibitem{LP01}
P.~Lahti and S.~Pulmannov{\'a}.
\newblock Coexistence vs.\ functional coexistence of quantum observables.
\newblock {\em Rep. Math. Phys.}, 47(2):199--212, 2001.

\end{thebibliography}
\end{document}